\pdfoutput=1

\documentclass[10pt,letter,prl,reprint,twocolumn,hypertext]{revtex4-1}

\usepackage{ifthen} 
\newboolean{pdflatex}
\setboolean{pdflatex}{true} 

\newboolean{articletitles}
\setboolean{articletitles}{true} 

\newboolean{uprightparticles}
\setboolean{uprightparticles}{false} 

\usepackage{microtype}
\usepackage{lineno}  
\usepackage{xspace} 
\usepackage{bigstrut}

\usepackage{graphicx}  
\usepackage{color}
\usepackage{colortbl}
\graphicspath{{./figs/}} 

\usepackage{amsmath} 
\usepackage{amssymb}
\usepackage{amsfonts}
\usepackage{upgreek} 

\newcommand*\patchAmsMathEnvironmentForLineno[1]{%
\expandafter\let\csname old#1\expandafter\endcsname\csname #1\endcsname
\expandafter\let\csname oldend#1\expandafter\endcsname\csname
end#1\endcsname
 \renewenvironment{#1}%
   {\linenomath\csname old#1\endcsname}%
   {\csname oldend#1\endcsname\endlinenomath}%
}
\newcommand*\patchBothAmsMathEnvironmentsForLineno[1]{%
  \patchAmsMathEnvironmentForLineno{#1}%
  \patchAmsMathEnvironmentForLineno{#1*}%
}
\AtBeginDocument{%
\patchBothAmsMathEnvironmentsForLineno{equation}%
\patchBothAmsMathEnvironmentsForLineno{align}%
\patchBothAmsMathEnvironmentsForLineno{flalign}%
\patchBothAmsMathEnvironmentsForLineno{alignat}%
\patchBothAmsMathEnvironmentsForLineno{gather}%
\patchBothAmsMathEnvironmentsForLineno{multline}%
}

\usepackage{hyperref}    
\usepackage[all]{hypcap} 
\usepackage{accents}    

\def\mup        {\ensuremath{\mu^+}\xspace}

\def\cquark    {\ensuremath{c}\xspace}
\def\squark    {\ensuremath{s}\xspace}
\def\bquark    {\ensuremath{b}\xspace}

\def\pion  {\ensuremath{\pi}\xspace}

\def\pip   {\ensuremath{\pion^+}\xspace}
\def\pim   {\ensuremath{\pion^-}\xspace}

\def\pipm  {\ensuremath{\pion^\pm}\xspace}
\def\pimp  {\ensuremath{\pion^\mp}\xspace}
\def\kaon  {\ensuremath{K}\xspace}
\def\Kbar  {\kern 0.2em\overline{\kern -0.2em K}{}\xspace}

\def\Kz    {\ensuremath{\kaon^0}\xspace}
\def\Kzb   {\ensuremath{\Kbar^0}\xspace}
\def\KzKzb {\ensuremath{\Kz \kern -0.16em \Kzb}\xspace}
\def\Kp    {\ensuremath{\kaon^+}\xspace}
\def\Km    {\ensuremath{\kaon^-}\xspace}
\def\Kpm   {\ensuremath{\kaon^\pm}\xspace}

\def\KpKm  {\ensuremath{\Kp \kern -0.16em \Km}\xspace}

\def\D       {\ensuremath{D}\xspace}
\def\Dz      {\ensuremath{\D^0}\xspace}

\def\Dsm     {\ensuremath{\D^-_\squark}\xspace}
\def\B       {\ensuremath{B}\xspace}
\def\Bbar    {\ensuremath{\kern 0.18em\overline{\kern -0.18em \B}{}}\xspace}

\def\Bd      {\ensuremath{\B^0}\xspace}
\def\Bs      {\ensuremath{\B^0_\squark}\xspace}
\def\Bsb     {\ensuremath{\Bbar^0_\squark}\xspace}
\def\Bdb     {\ensuremath{\Bbar^0}\xspace}
\def\Bu      {\ensuremath{\B^+}\xspace}
\def\Bub     {\ensuremath{\B^-}\xspace}
\def\Bp      {\ensuremath{\Bu}\xspace}
\def\Bm      {\ensuremath{\Bub}\xspace}
\def\Bpm     {\ensuremath{\B^\pm}\xspace}

\def\jpsi     {\ensuremath{{J\mskip -3mu/\mskip -2mu\psi\mskip 2mu}}\xspace}
\def\proton      {\ensuremath{p}\xspace}
\def\antiproton  {\ensuremath{\overline \proton}\xspace}
\def\Lbar {\ensuremath{\kern 0.1em\overline{\kern -0.1em\Lambda}}\xspace}

\def\CP                {\ensuremath{C\!P}\xspace}
\def\CPT               {\ensuremath{C\!PT}\xspace}
\newcommand{\stat}{\ensuremath{\mathrm{\,(stat)}}\xspace}
\newcommand{\syst}{\ensuremath{\mathrm{\,(syst)}}\xspace}
\def\lhcb {\mbox{LHCb}\xspace}
\def\pt         {\mbox{$p_{\rm T}$}\xspace}
\newcommand{\chisq}{\ensuremath{\chi^2}\xspace}

\def\pythia     {\mbox{\textsc{Pythia}}\xspace}
\def\evtgen     {\mbox{\textsc{EvtGen}}\xspace}
\def\photos     {\mbox{\textsc{Photos}}\xspace}
\def\geant      {\mbox{\textsc{Geant4}}\xspace}
\def\gauss      {\mbox{\textsc{Gauss}}\xspace}

\newcommand{\tev}{\ifthenelse{\boolean{inbibliography}}{\ensuremath{~T\kern -0.05em eV}\xspace}{\ensuremath{\mathrm{\,Te\kern -0.1em V}}\xspace}}
\newcommand{\gev}{\ensuremath{\mathrm{\,Ge\kern -0.1em V}}\xspace}
\newcommand{\mev}{\ensuremath{\mathrm{\,Me\kern -0.1em V}}\xspace}
\newcommand{\kev}{\ensuremath{\mathrm{\,ke\kern -0.1em V}}\xspace}
\newcommand{\ev}{\ensuremath{\mathrm{\,e\kern -0.1em V}}\xspace}
\newcommand{\gevc}{\ensuremath{{\mathrm{\,Ge\kern -0.1em V\!/}c}}\xspace}
\newcommand{\mevc}{\ensuremath{{\mathrm{\,Me\kern -0.1em V\!/}c}}\xspace}
\newcommand{\gevcc}{\ensuremath{{\mathrm{\,Ge\kern -0.1em V\!/}c^2}}\xspace}
\newcommand{\gevgevcccc}{\ensuremath{{\mathrm{\,Ge\kern -0.1em V^2\!/}c^4}}\xspace}
\newcommand{\mevcc}{\ensuremath{{\mathrm{\,Me\kern -0.1em V\!/}c^2}}\xspace}

\def\pipipi {\ensuremath{{\Bpm \to \pip \pim \pipm}}\xspace}
\def\kpipi {\ensuremath{{\Bpm \to \Kpm \pip \pim}}\xspace}
\def\kkpi {\ensuremath{{\Bpm \to \Kp \Km \pipm}}\xspace}
\def\kkk {\ensuremath{{\Bpm \to \Kpm \Kp \Km}}\xspace}
\def\pppi {\ensuremath{{\Bpm \to p \bar{p} \pi^{\pm}}}\xspace}
\def\ppk {\ensuremath{{\Bpm \to p \bar{p} K^{\pm}}}\xspace}
\def\jpsik {\ensuremath{{\Bpm \to \jpsi \Kpm}}\xspace}

\def\mpipi {\ensuremath{m_{\pip \pim}}\xspace}                        
\def\mmkk {\ensuremath{m^2_{\Kp \Km}}\xspace}                            
\def\mkk {\ensuremath{m_{\Kp \Km}}\xspace}                            
\def\mkpi {\ensuremath{m_{\Kpm \pimp}}\xspace}                            
\def\mmpipilow {\ensuremath{m^2_{\pip \pim\,{\rm low}}}\xspace} 
\def\mmpipihi {\ensuremath{m^2_{\pip \pim\,{\rm high}}}\xspace}   

\def\acp {\ensuremath{A_{\CP}}\xspace}
\def\acpraw {\ensuremath{A_{\rm raw}}\xspace} 

\def\acpn {\ensuremath{A_{\rm raw}^{N}}\xspace}
\def\adetk {\ensuremath{A_{\rm D}(\Kpm)}\xspace}
\def\aprod {\ensuremath{A_{\rm P}(\Bpm)}\xspace}
\def\adetpi {\ensuremath{A_{\rm D}(\pipm)}\xspace}

\usepackage{mciteplus}

\begin{document}



\title{Measurement of $C\!P$ violation in the phase space of\\ $B^{\pm} \rightarrow K^{+} K^{-} \pi^{\pm}$ and $B^{\pm} \rightarrow \pi^{+} \pi^{-} \pi^{\pm}$ decays \vspace{0.4cm}}



\begin{abstract}
The charmless decays $B^{\pm} \rightarrow K^{+}K^{-}\pi^{\pm}$ and $B^{\pm} \rightarrow \pi^{+}\pi^{-}\pi^{\pm}$ are reconstructed in a data set, corresponding to an integrated luminosity of 1.0~fb$^{-1}$ of $pp$ collisions at a center-of-mass energy of 7~TeV, collected by LHCb in 2011. 
The inclusive charge asymmetries of these modes are measured to be 
$ A_{C\!P}(B^{\pm} \rightarrow K^{+}K^{-}\pi^{\pm}) =-0.141 \pm 0.040 \stat \pm 0.018 \syst \pm 0.007 (\jpsi K^{\pm})$ and $A_{C\!P}(B^{\pm} \rightarrow \pi^{+}\pi^{-}\pi^{\pm}) = 0.117 \pm 0.021 \stat \pm 0.009 \syst \pm 0.007 (\jpsi K^{\pm})$, 
where the third uncertainty is due to the $C\!P$ asymmetry of the \jpsik reference mode. 
In addition to the inclusive $C\!P$ asymmetries, larger asymmetries are observed in localized regions of phase space. 
\end{abstract}

\vspace*{1.cm}
\author{
\centerline{\large\bf LHCb collaboration}
\begin{flushleft}
\small
R.~Aaij$^{40}$, 
B.~Adeva$^{36}$, 
M.~Adinolfi$^{45}$, 
C.~Adrover$^{6}$, 
A.~Affolder$^{51}$, 
Z.~Ajaltouni$^{5}$, 
J.~Albrecht$^{9}$, 
F.~Alessio$^{37}$, 
M.~Alexander$^{50}$, 
S.~Ali$^{40}$
% ...

{\footnotesize \it
$ ^{1}$Centro Brasileiro de Pesquisas F\'{i}sicas (CBPF), Rio de Janeiro, Brazil\\
% ...
$ ^{5}$Clermont Universit\'{e}, Universit\'{e} Blaise Pascal, CNRS/IN2P3, LPC, Clermont-Ferrand, France\\
$ ^{6}$CPPM, Aix-Marseille Universit\'{e}, CNRS/IN2P3, Marseille, France\\
% ...
$ ^{9}$Fakult\"{a}t Physik, Technische Universit\"{a}t Dortmund, Dortmund, Germany\\
% ...
$ ^{36}$Universidad de Santiago de Compostela, Santiago de Compostela, Spain\\
$ ^{37}$European Organization for Nuclear Research (CERN), Geneva, Switzerland\\
% ...
$ ^{40}$Nikhef National Institute for Subatomic Physics, Amsterdam, The Netherlands\\
% ...
$ ^{45}$H.H. Wills Physics Laboratory, University of Bristol, Bristol, United Kingdom\\
% ...
$ ^{50}$School of Physics and Astronomy, University of Glasgow, Glasgow, United Kingdom\\
$ ^{51}$Oliver Lodge Laboratory, University of Liverpool, Liverpool, United Kingdom\\
% ...
}
\end{flushleft}
}
\collaboration{The LHCb collaboration\vspace{0.4cm}}

\pacs{13.25.Hw,11.30.Er}

\vspace*{-1.5cm}
\hspace{-8.6cm}
\mbox{\Large EUROPEAN ORGANIZATION FOR NUCLEAR RESEARCH (CERN)}

\vspace*{0.7cm}
\hspace*{-9cm}
\begin{tabular*}{16.6cm}{lc@{\extracolsep{\fill}}r}
\ifthenelse{\boolean{pdflatex}}
{\vspace*{-3.2cm}\mbox{\!\!\!\includegraphics[width=.14\textwidth]{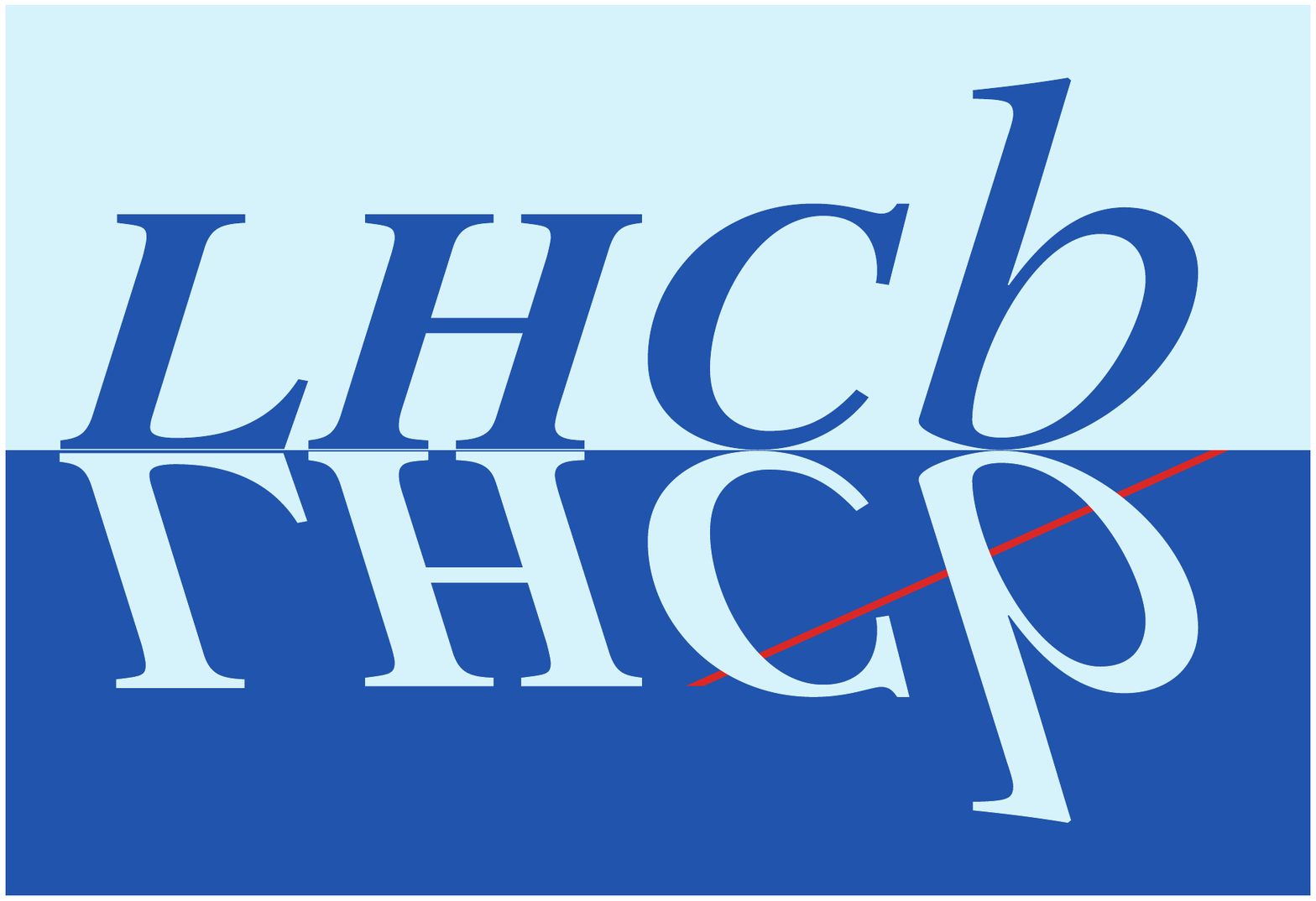}} & &}%
{\vspace*{-1.2cm}\mbox{\!\!\!\includegraphics[width=.12\textwidth]{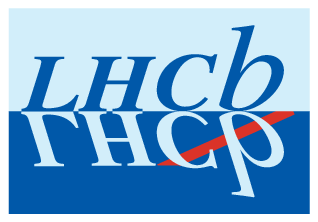}} & &}%
\\ 
 & & CERN-PH-EP-2013-190 \\  
 & & LHCb-PAPER-2013-051 \\  
 & &17 October 2013 \\ 
 & & \\
\end{tabular*}

\vspace*{21.0cm}
\hspace*{-5cm}\centerline{\copyright~CERN on behalf of the LHCb collaboration, license \href{http://creativecommons.org/licenses/by/3.0/}{CC-BY-3.0}.}
\vspace*{-21.0cm}

\vspace*{20.0cm}
\hspace*{-5cm}\centerline{\large Submitted to Phys.~Rev.~Lett.}
\vspace*{-20.0cm}


\boldmath
\maketitle
\unboldmath



Charmless decays of $B$ mesons to three hadrons are dominated by quasi-two-body processes involving intermediate resonant states. 
The rich interference pattern present in such decays makes them favorable for the investigation of charge asymmetries that are localized in the phase space~\cite{Miranda1,Miranda2}.
The large samples of charmless $B$ decays collected by the LHCb experiment allow direct \CP violation to be measured in regions of phase space.   
In previous measurements of this type, the phase spaces of \kkk and \kpipi decays were observed to have regions of large local asymmetries~\cite{LHCb-PAPER-2013-027}. 
Concerning baryonic modes, no significant effects have been observed  in either \ppk or \pppi decays~\cite{LHCB-PAPER-2013-031}. 
Large \CP-violating asymmetries have also been observed in charmless two-body $B$ meson decays such as $\Bd \to\Kp\pim$ and $\Bs\to\Km\pip$ (and the corresponding \Bdb and \Bsb decays)~\cite{LHCb-PAPER-2013-018}.

Some recent efforts have been made to understand the origin of the large asymmetries. 
For direct \CP violation to occur, two interfering amplitudes with different weak and strong phases must be involved in the decay process~\cite{BSS1979}.  
Interference between intermediate states of the decay can introduce large strong phase differences, and is one mechanism for explaining local asymmetries in the phase space~\cite{PhysRevD.87.076007,Bhattacharya:2013cvn}.
Another explanation focuses on final-state $KK \leftrightarrow \pi \pi$ rescattering, which can occur between decay channels with the same flavor quantum numbers~\cite{LHCb-PAPER-2013-027,Bhattacharya:2013cvn,IgnacioCPT}. 
Invariance of \CPT symmetry constrains hadron rescattering so that the sum of the partial decay widths, for all channels with the same final-state quantum numbers related by the S matrix, must be equal for charge-conjugated decays.  
Effects of SU(3) flavor symmetry breaking have also been investigated and partially explain the observed patterns~\cite{Xu:2013dta,Bhattacharya:2013cvn,Gronau:2013mda}. 

The \kkpi decay is interesting because $s\bar{s}$ resonant contributions are strongly suppressed~\cite{ozzi1,ozzi2,ozzi4}. 
Recently, LHCb reported an upper limit on the $\phi$ contribution to be $\mathcal{B}(\Bpm \to \phi\pipm) < 1.5\times 10^{-7} $ at the 90\% confidence level~\cite{LHCB-PAPER-2013-048}. 
The lack of $\Kp\Km$ resonant contributions makes the \kkpi decay a good probe for rescattering from decays with pions. 
The \pipipi mode, on the other hand, has large resonant contributions, as shown in an amplitude analysis by the BaBar collaboration, which measured the inclusive \CP asymmetry to be $(0.03 \pm 0.06)$~\cite{BaBarpipipi}. 
For \kkpi decays, the inclusive \CP-violating asymmetry was measured by the BaBar collaboration to be ($0.00 \pm 0.10$)~\cite{BaBarkkpi}, from a comparison of \Bp and \Bm sample fits. 
Both results are compatible with the no \CP-violation hypothesis.

In this Letter we report  measurements of the inclusive \CP-violating asymmetries for \pipipi and \kkpi decays. 
The \CP asymmetry in \Bpm decays to a final state $f^{\pm}$ is defined as
\begin{equation}
\acp(\Bpm \to f^{\pm}) \equiv \Phi [ \Gamma(\Bm \to  f^{-}) , \Gamma(\Bp \to  f^{+})],
\end{equation}
where $\Phi[X,Y]\equiv (X-Y)/(X+Y)$ is the asymmetry function, 
$\Gamma$ is the decay width, and the final states $f^{\pm}$ are $\pip \pim\pipm$ or $\Kp \Km\pipm$.
The asymmetry distributions across the phase space are also investigated.

The \lhcb detector~\cite{Alves:2008zz} is a single-arm forward spectrometer covering the \mbox{pseudorapidity} range $2<\eta <5$, designed for the study of particles 
containing \bquark or \cquark quarks. The analysis is based on $pp$ collision data, corresponding to an integrated luminosity of 1.0\,fb$^{-1}$, collected in 2011 at 
a center-of-mass energy of 7~TeV.

\begin{figure*}[tb]
\centering
\includegraphics[width=0.48\linewidth]{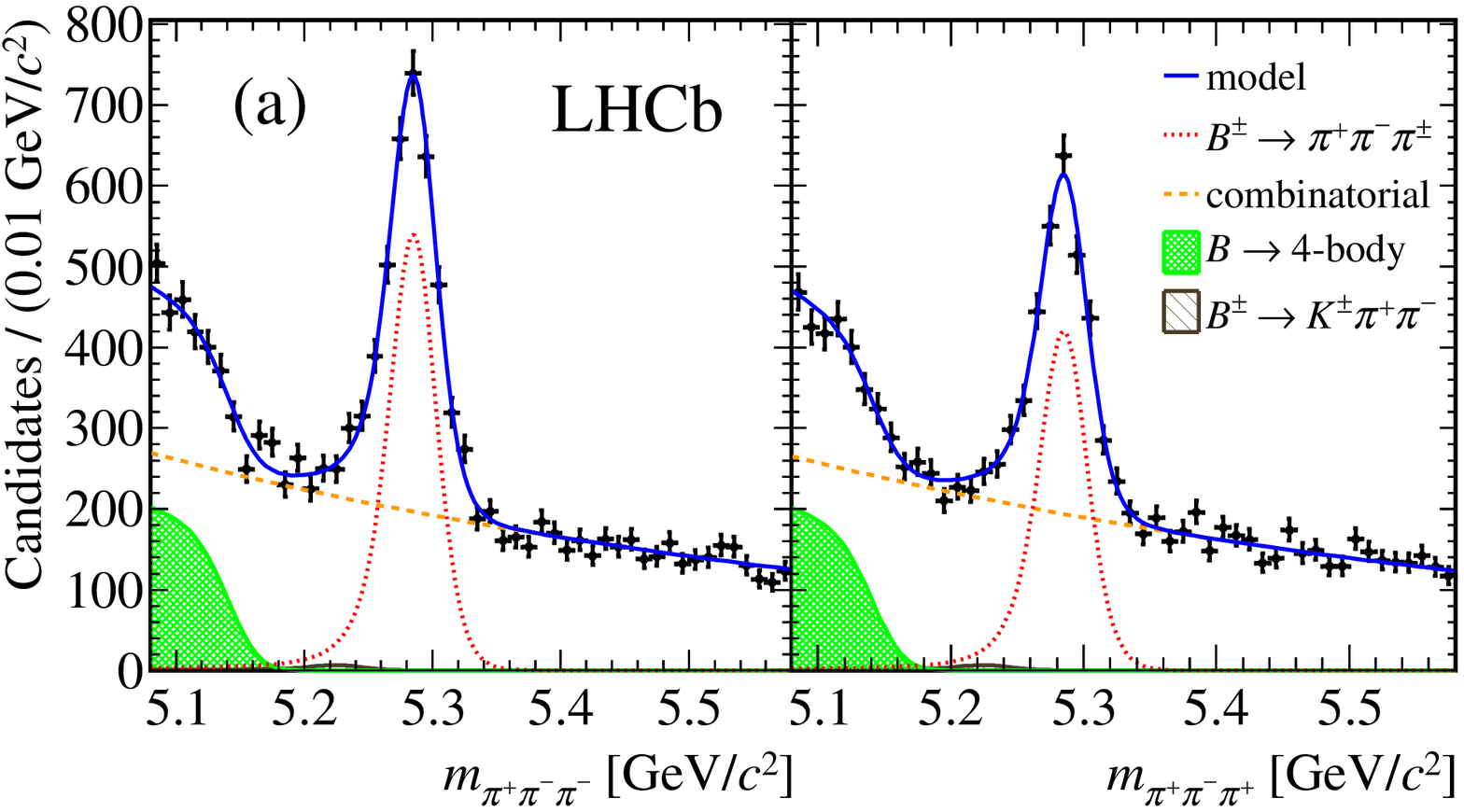}
\hspace{0.2cm}
\includegraphics[width=0.48\linewidth]{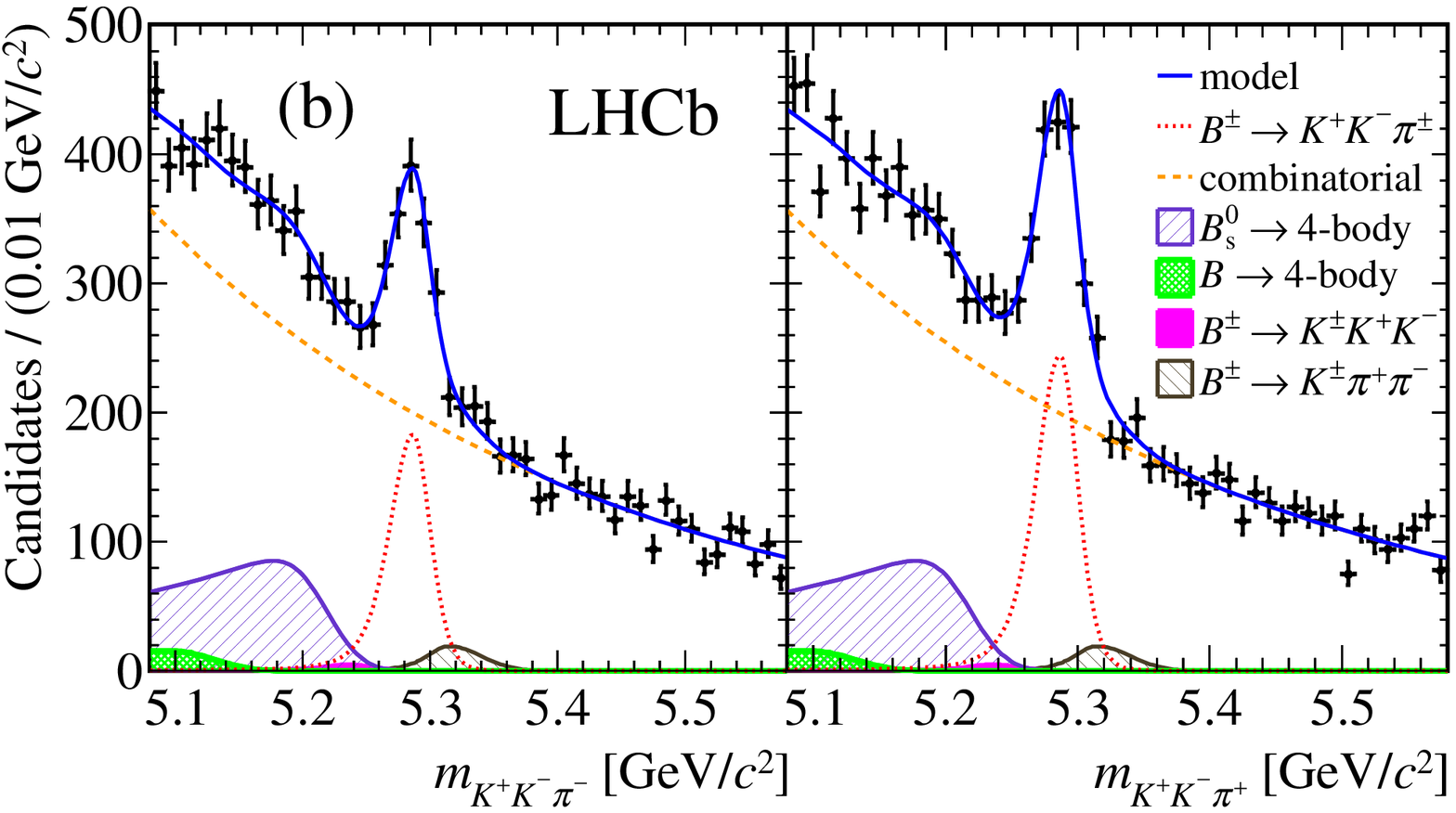}
\caption{Invariant mass spectra of (a) \pipipi decays and (b) \kkpi decays. The left panel in each figure shows the \Bm modes and the right panel shows the \Bp modes.
The results of the unbinned maximum likelihood fits are overlaid. The main components of the fit are also shown. 
}
\label{MassFit}
\end{figure*}

Events are selected by a trigger~\cite{LHCb-DP-2012-004} that consists of a hardware stage, based on information from a calorimeter system and five muon stations, followed by a software stage, which applies a full event reconstruction. 
Candidate events are first required to pass the hardware trigger, which selects particles with a large 
transverse energy. 
The software trigger requires a two-, three- or four-track secondary vertex with a high sum of the transverse momenta, \pt, of the tracks and significant displacement from the primary $pp$ interaction vertices~(PVs). 
At least one track should have $\pt > 1.7\gevc$ and $\chisq_{\rm IP}$ with respect to any 
primary interaction greater than 16, where $\chisq_{\rm IP}$ is defined as the difference between the \chisq of a given PV reconstructed with and without the considered track, and IP is the impact parameter. 
A multivariate algorithm~\cite{BBDT} is used for the identification of secondary vertices consistent with the decay of a \bquark hadron.

Further criteria are applied offline to select $B$ mesons and suppress the combinatorial background. 
The \Bpm decay products are required to satisfy a set of selection criteria on their momenta, their \pt, the $\chisq_{\rm IP}$ of the final-state tracks, and the distance of closest approach between any two tracks. 
The $B$ candidates are required to have $\pt > 1.7\gevc$, $\chisq_{\rm IP}<10$ (defined by projecting the $B$ candidate trajectory backwards from its decay vertex), decay vertex $\chisq<12$, and  decay vertex displacement from any PV greater than 3~mm. 
Additional requirements are applied to variables related to the $B$-meson production and decay, such as the angle $\theta$ between the $B$-candidate momentum and the direction of flight from the primary vertex to the decay vertex, $\cos(\theta)>0.99998$. 
Final-state kaons and pions are further selected using particle identification information, provided by two ring-imaging Cherenkov detectors~\cite{LHCb-DP-2012-003}, and are required to be incompatible with a muon~\cite{LHCb-DP-2013-001}. 
The kinematic selection is 
common to both decay channels, while the particle identification selection is specific to each final state. Charm contributions are removed by excluding the regions 
of $\pm 30 \mevcc$ around the world average value of the $\Dz$ mass~\cite{PDG2012} in the two-body invariant masses \mpipi, \mkpi and \mkk. 

The simulated events used in this analysis are generated 
using \pythia~6.4~\cite{Sjostrand:2006za} with a specific \lhcb configuration~\cite{LHCb-PROC-2010-056}.  
Decays of hadronic particles are produced by \evtgen~\cite{Lange:2001uf}, in which final-state radiation is generated using \photos~\cite{Golonka:2005pn}. 
The interaction of the generated particles with the detector and its response are implemented using the \geant toolkit~\cite{Allison:2006ve, *Agostinelli:2002hh} as 
described in Ref.~\cite{LHCb-PROC-2011-006}.

Unbinned extended maximum likelihood fits to the mass spectra of the selected \Bpm candidates are performed to obtain the signal yields and raw asymmetries. 
The \kkpi and \pipipi signal components are parametrized by a {\mbox{Cruijff}} function~\cite{Cruijff} with equal left and right widths and different radiative tails to account for the asymmetric effect of final-state radiation on the signal shape. 
The means and widths are left to float in the fit, while the tail parameters are fixed to the values obtained from simulation. 
The combinatorial background is described by an exponential distribution whose parameter is left free in the fit.
The backgrounds due to partially reconstructed four-body \B decays are parametrized by an ARGUS distribution~\cite{Argus} convolved with a Gaussian resolution function.
For \pipipi decays the shape and yield parameters describing the backgrounds are varied in the fit, while for \kkpi decays they are taken from simulation, due to a further contribution from four-body \Bs decays such as $\Bs \to \Dsm (\Kp \Km \pim) \pip$. 
We define peaking backgrounds as decay modes with one misidentified particle, namely the channels \kpipi for the \pipipi mode, and \kpipi and \kkk for the \kkpi mode. 
The shapes and yields of the peaking backgrounds  are obtained from simulation. 
The yields of the peaking and partially reconstructed background components are constrained to be equal for \Bp and \Bm decays. 
The invariant mass spectra of the \kkpi and \pipipi candidates are shown in Fig.~\ref{MassFit}.

The signal yields obtained are $N(KK\pi)=1870\pm133$ and $N(\pi\pi\pi)=4904\pm 148$, and the raw asymmetries are
$\acpraw(K\!K\pi)=-0.143\pm 0.040$ and $\acpraw(\pi\pi\pi) = 0.124\pm 0.020$, where the uncertainties are statistical. The \CP asymmetries are expressed in terms of the 
measured raw asymmetries, corrected for effects induced by the detector acceptance and interactions of final-state pions with matter \adetpi, as well as for a 
possible $\B$-meson production asymmetry \aprod, 
\begin{equation}
\!\!\! \acp \!=\! \acpraw \! -\! \adetpi \!-\! \aprod .
\label{eq:acpsum}
\end{equation}
The pion detection asymmetry, $\adetpi =0.0000\pm0.0025$, has been previously measured by LHCb~\cite{LHCb-PAPER-2012-009}.
The production asymmetry \aprod is measured from a data sample of approximately $6.3 \times 10^4$ $\Bpm \to \jpsi (\mup\mu^-)\Kpm$ decays. 
The \jpsik sample passes the same trigger, kinematic, and kaon particle identification selection criteria as the signal samples, and it has a similar event topology. 
The \aprod term is obtained from the raw asymmetry of the \jpsik mode as 
\begin{equation}
\aprod = \acpraw(\jpsi K) - \acp(\jpsi K) - \adetk, 
\label{deltaJpsik}
\end{equation}
where $\acp(\jpsi K) = 0.001\pm 0.007$~\cite{PDG2012} is the world average \CP asymmetry of \jpsik decays, and $\adetk =-0.010\pm0.003$ is the kaon interaction asymmetry obtained from $\Dz\to \Kpm\pimp$ and $\Dz\to \Kp\Km$ decays~\cite{LHCb-PAPER-2011-029}, and corrected for \adetpi. 
The \CP asymmetries of the \kkpi and \pipipi channels are then determined using Eqs.~\ref{eq:acpsum} and~\ref{deltaJpsik}.

\begin{figure*}[tb]
\centering
\includegraphics[width=0.49\linewidth]{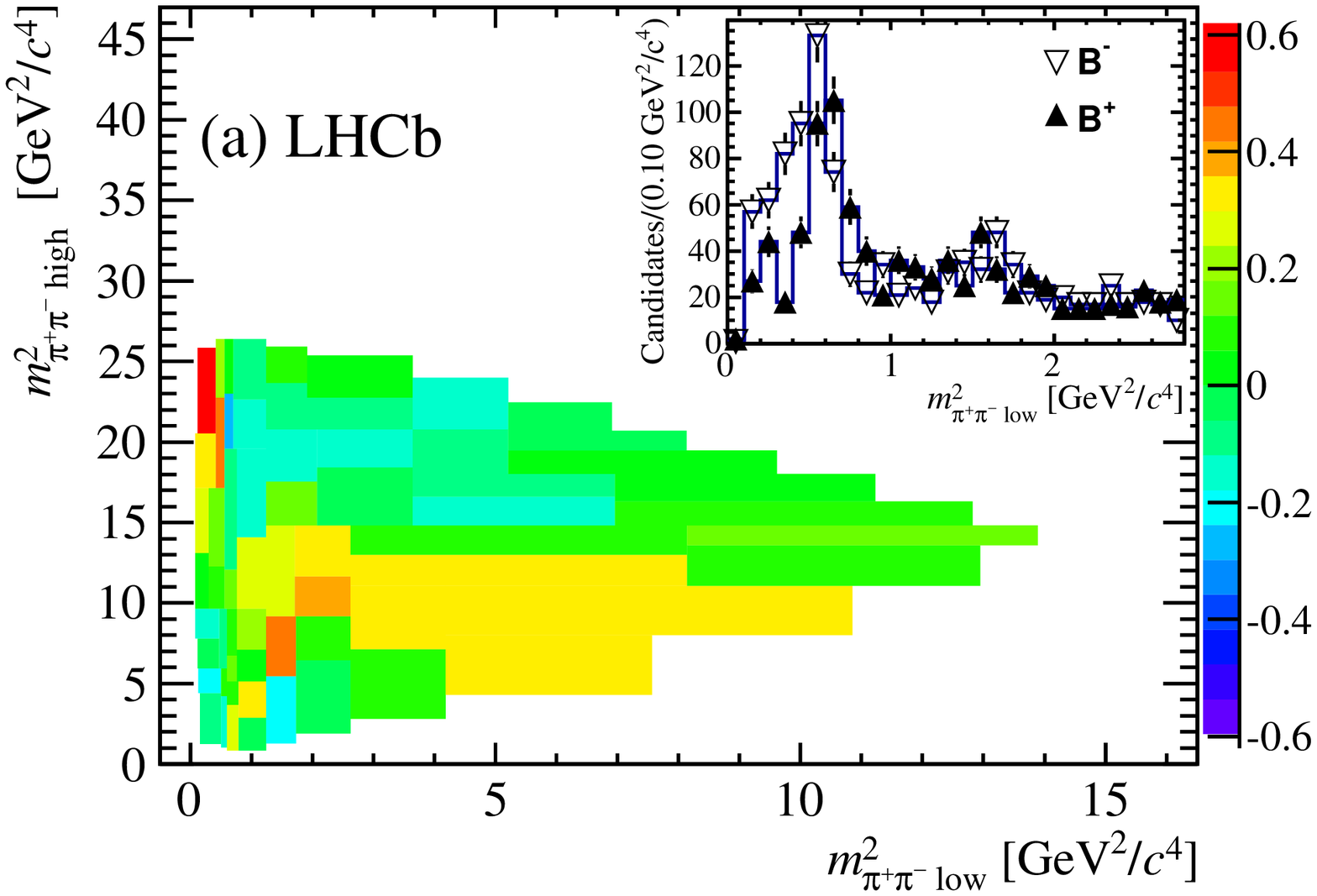}
\includegraphics[width=0.49\linewidth]{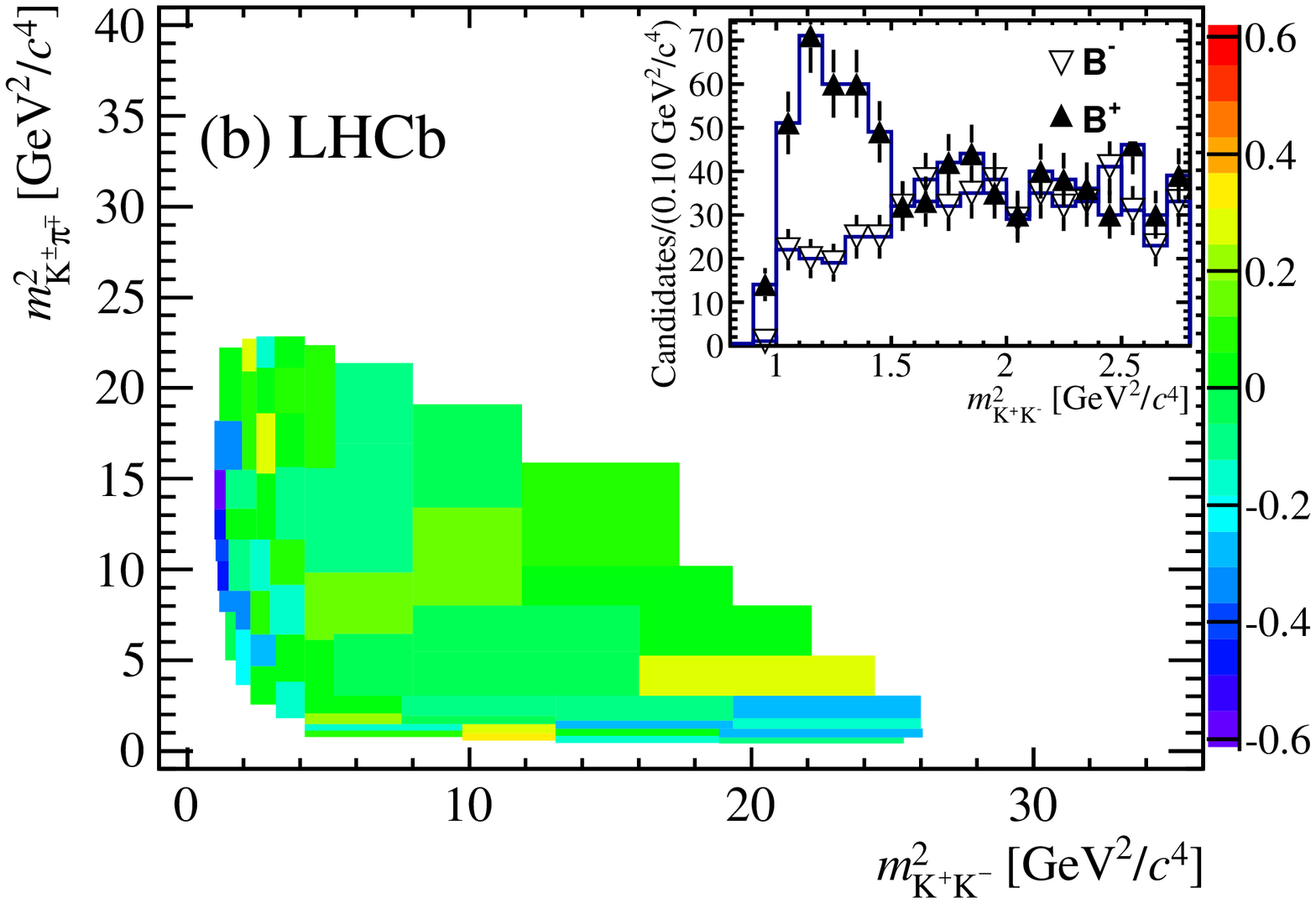}
\caption{
Asymmetries of the number of events (including signal and background) in bins of the Dalitz plot, \acpn, for (a) \pipipi and  (b)~\kkpi decays.  
The inset figures show the projections of the number of events in bins of  (a) the \mmpipilow variable 
for $\mmpipihi>15\gevgevcccc$ and (b) the \mmkk variable. 
The distributions are not corrected for efficiency.
}
\label{Mirandizing}
\end{figure*}

Since the detector efficiencies for the signal modes are not uniform across the Dalitz plot, and the raw asymmetries are also not uniformly distributed, an acceptance correction is applied to the integrated raw asymmetries. 
It is determined by the ratio between the \Bm and \Bp average efficiencies in simulated events, reweighted to reproduce the population of signal data over the phase space. 
Furthermore, the detector acceptance and reconstruction depend on the trigger selection. 
The efficiency of the hadronic hardware trigger is found to have a small charge asymmetry for kaons. 
Therefore, the data are divided into two samples: events with candidates selected by the hadronic trigger and events selected by other triggers independently of the signal candidate. 
The acceptance correction and subtraction of the \aprod term is performed separately for each trigger configuration. 
The trigger-averaged value of the production asymmetry is $\aprod=-0.004\pm0.004$, where the uncertainty is statistical only. 
The integrated \CP asymmetries are then the weighted averages of the \CP asymmetries for the two trigger samples.

The methods used in estimating the systematic uncertainties of the signal model, combinatorial background, peaking background, and acceptance correction are the same as those used in Ref.~\cite{LHCb-PAPER-2013-027}. 
For \kkpi decays, we also evaluate a systematic uncertainty due to the partially reconstructed background model by varying the mean and resolution according to the difference between simulation and data obtained from the signal component. 
The \adetpi and \adetk uncertainties are included as systematic uncertainties related to the procedure. 
A systematic uncertainty is also evaluated to account for the difference in kaon kinematics between the \Bpm and \Dz decays. 
The systematic uncertainties for the measurements of $\acp(\kkpi)$ and $\acp(\pipipi)$ are summarized in Table~\ref{tab:syst}.

The results obtained for the inclusive \CP asymmetries of the \kkpi and \pipipi decays are
\begin{eqnarray}
 \acp( \kkpi)  \! &=& \!-0.141 \pm  0.040  \pm  0.018 \pm 0.007  ,  \nonumber  \\ [1mm]
 \acp(\pipipi)  \! & = &\!  0.117 \pm 0.021  \pm  0.009 \pm 0.007  ,  \nonumber
\end{eqnarray}
where the first uncertainty is statistical, the second is the experimental systematic, and the third is due to the \CP asymmetry of the \jpsik reference 
mode~\cite{PDG2012}. The significances of the inclusive charge asymmetries, calculated by dividing the central values by the sum in quadrature of the statistical and 
both systematic uncertainties, are 3.2 standard deviations~($\sigma$) for \kkpi and $4.9\sigma$ for \pipipi decays.

\begin{table}[b]
    \small
  \caption{    \small
   Systematic uncertainties on $\acp(\kkpi)$ and $\acp(\pipipi)$. The total systematic uncertainties are the sum in quadrature of the individual 
contributions.}
\begin{center}\begin{tabular}{lcc}
\hline \hline
   Systematic uncertainty & $\acp(K\!K\pi)$ & $\acp(\pi\pi\pi)$      \\ 
    \hline
Signal model              & \;\;\;\,0.001  & \;\;\;\,0.0005  \\
Combinatorial background  & \;\;\;\,0.003  & \;\;\;\,0.0008 \\
Peaking background        & $\;\;\;\,0.001$& $\;\;\;\,0.0025$\\
Acceptance                & \;\;\;\,0.014 & \;\;\;\,0.0032 \\ 
Part. rec. background   & \;\;\;\,0.005  & \;\;\;\,--  \\ 
\adetpi uncertainty  & \;\;\;\,0.003 & \;\;\;\,0.0025 \\ 
\adetk uncertainty   & \;\;\;\,0.003  & \;\;\;\,0.0032  \\ 
\adetk kaon kinematics    & \;\;\;\,0.008 & \;\;\;\,0.0075 \\ 
\hline
Total                     & \;\;\;\,0.018  & \;\;\;\,0.0094   \\
\hline \hline
  \end{tabular}\end{center}
\label{tab:syst}
\end{table}

\begin{figure*}[!tb]
\centering
\includegraphics[width=0.48\linewidth]{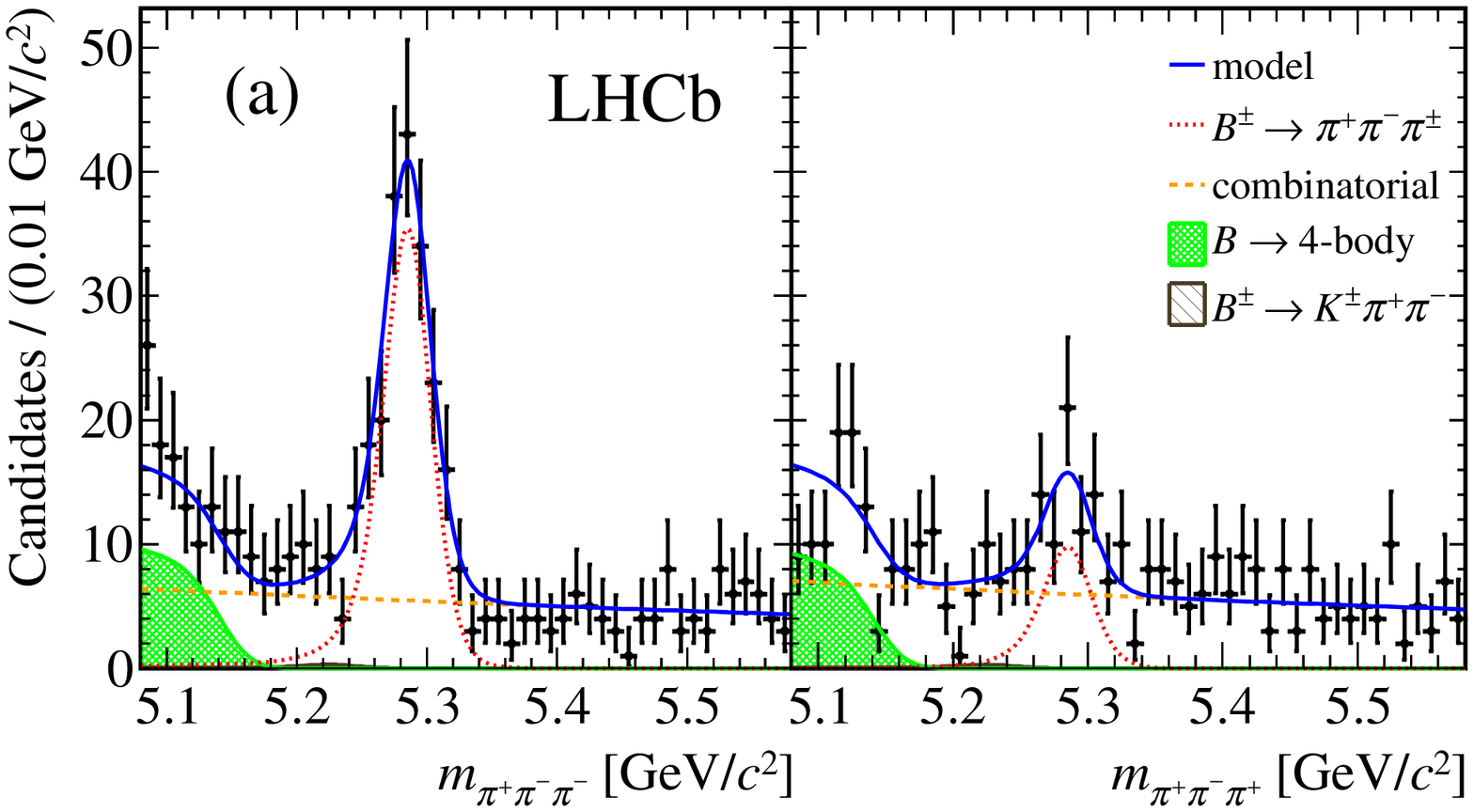}
\hspace{0.2cm}
\includegraphics[width=0.48\linewidth]{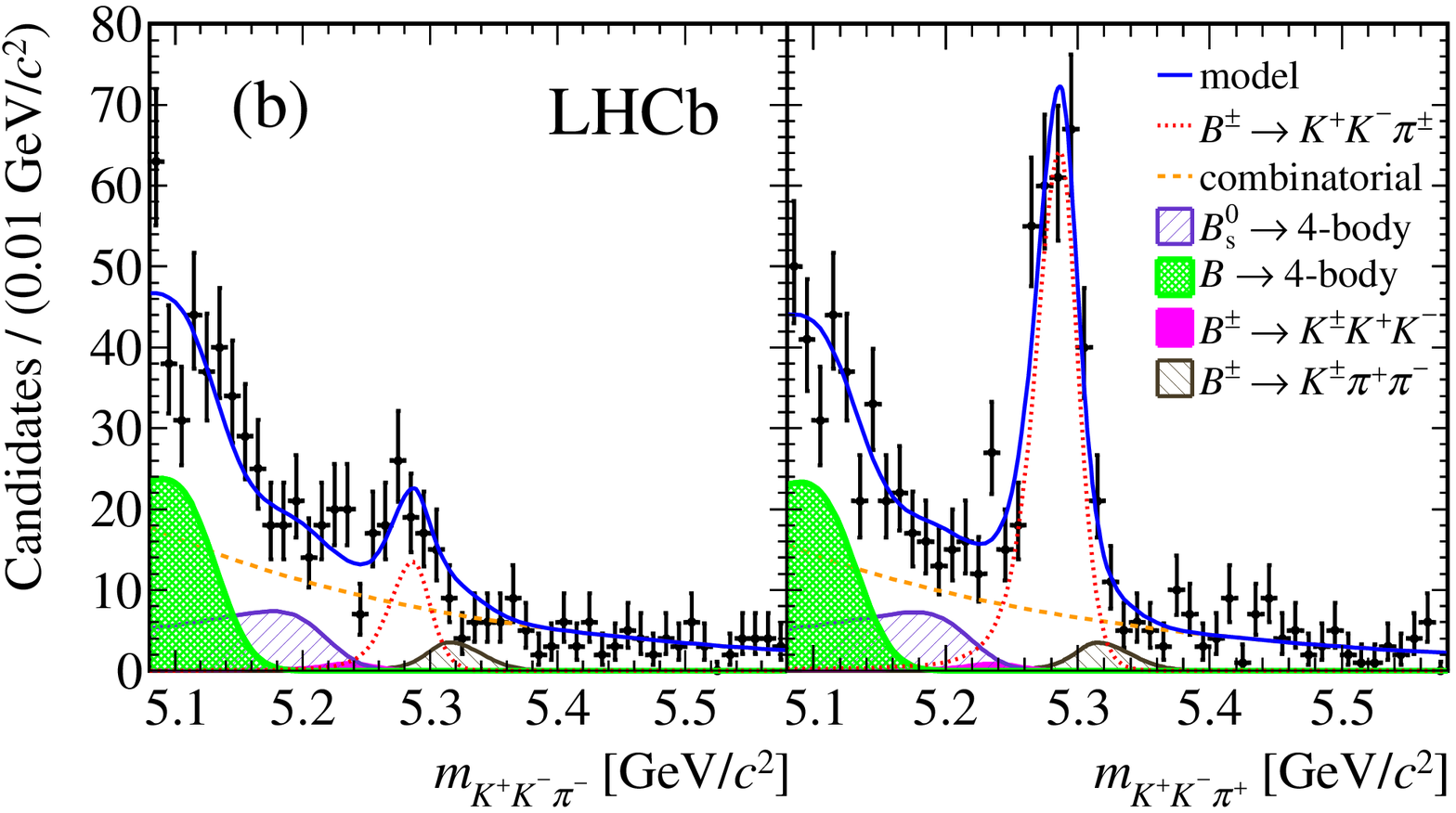}
\caption{Invariant mass spectra of (a) \pipipi decays in the region $\mmpipilow < 0.4\gevgevcccc$ and $\mmpipihi > 15\gevgevcccc$, and  (b) \kkpi decays in the 
region $\mmkk < 1.5\gevgevcccc$. The left panel in each figure shows the \Bm modes and the right panel shows the \Bp modes. The results of the unbinned maximum 
likelihood fits are overlaid. 
}
\label{MassFitRegion}
\end{figure*}

In addition to the inclusive charge asymmetries, we study the asymmetry distributions in the two-dimensional phase space of two-body invariant masses. 
The Dalitz plot distributions in the signal region, defined as the three-body invariant mass region within two Gaussian widths from the signal peak, are divided into bins with approximately equal numbers of events in the combined \Bm and \Bp samples. 
Figure~\ref{Mirandizing} shows the raw asymmetries (not corrected for efficiency), $\acpn =  \Phi [N^{-}, N^{+}]$, computed using the number of negative ($N^{-}$) and positive ($N^{+}$) entries in each bin of the \pipipi and \kkpi Dalitz plots.
The \pipipi Dalitz plot is symmetrized and its two-body invariant mass squared variables are defined as $\mmpipilow < \mmpipihi$. 
The \acpn distribution in the Dalitz plot of the \pipipi mode reveals an asymmetry concentrated at low values of \mmpipilow and high values of \mmpipihi. 
The distribution of the projection of the number of events onto the \mmpipilow invariant mass (inset in Fig.~\ref{Mirandizing}(a)) shows that this asymmetry is located in the region $\mmpipilow < 0.4\gevgevcccc$ and $\mmpipihi >15\gevgevcccc$.  
For \kkpi we identify a negative asymmetry located in the low $\Kp\Km$ invariant mass region. 
This can be seen also in the inset figure of the $\Kp\Km$ invariant mass projection, where there is an excess of \Bp candidates for $\mmkk < 1.5\gevgevcccc$. 
Although \kkpi has no $\phi(1020)$ contribution~\cite{LHCB-PAPER-2013-048,phiBR}, a clear structure  is observed. 
This structure was also seen by the BaBar collaboration~\cite{BaBarkkpi} but was not studied separately for \Bm and \Bp components. 
No significant asymmetry is present in the low-mass region of the ${\Kpm \pimp}$ invariant mass projection.

The \CP asymmetries are further studied in the regions where large raw asymmetries are found. 
The regions are defined as $\mmpipihi > 15\gevgevcccc$ and $\mmpipilow < 0.4\gevgevcccc$ for the \pipipi mode, and $\mmkk < 1.5\gevgevcccc$ for the \kkpi mode.
Unbinned extended maximum likelihood fits are performed to the mass spectra of the candidates in these regions, using the same models as for the global fits. 
The spectra are shown in Fig.~\ref{MassFitRegion}. 
The resulting signal yields and raw asymmetries for the two regions are ${N^{\mathrm {reg}}(K\!K\pi)=342\pm28}$ and 
${\acpraw^{\mathrm {reg}}(K\!K\pi)=-0.658\pm0.070}$ for the \kkpi mode, and ${N^{\mathrm {reg}}(\pi\pi\pi)=229\pm20}$ and 
${\acpraw^{\mathrm {reg}}(\pi\pi\pi)=0.555\pm0.082}$ for the \pipipi mode. The \CP asymmetries are obtained from the raw asymmetries using Eqs.~\ref{eq:acpsum} and~\ref{deltaJpsik} and applying an acceptance correction. 
Systematic uncertainties are estimated due to the signal models, acceptance correction and binning choice in the region, the \adetpi and \aprod statistical uncertainties and the \adetk kaon kinematics. 
The local charge asymmetries for the two regions are measured to be
\begin{eqnarray}
\acp^{\mathrm {reg}}(\kkpi) \!  &=&\! -0.648 \pm 0.070 \pm 0.013 \pm 0.007,  \nonumber  \\ [1mm]
\acp^{\mathrm {reg}}(\pipipi)  \! &=& \! 0.584  \pm 0.082 \pm 0.027 \pm 0.007 , \nonumber
\end{eqnarray}
where the first uncertainty is statistical, the second is the experimental systematic and the third is due to the \CP asymmetry of the \jpsik reference mode~\cite{PDG2012}.

In conclusion, we have found the first evidence of inclusive \CP asymmetries of the \kkpi and \pipipi 
modes with significances of $3.2\sigma$ and $4.9\sigma$, respectively. 
The results are consistent with those measured by the BaBar collaboration~\cite{BaBarkkpi,BaBarpipipi}.
These charge asymmetries are not uniformly distributed in the phase space. 
For  \kkpi decays, where no significant resonant contribution is expected, we observe 
a very large negative asymmetry  concentrated in a restricted region of the phase space in 
the  low $\Kp\Km$ invariant mass. 
For \pipipi decays, a large positive asymmetry is measured in the low \mmpipilow and high \mmpipihi phase-space region, not clearly associated to a resonant state.
The evidence presented here for  \CP violation in \kkpi and \pipipi decays, along with the recent evidence for \CP violation in \kpipi and \kkk decays~\cite{LHCb-PAPER-2013-027} and recent theoretical developments~\cite{Bhattacharya:2013cvn,IgnacioCPT,Xu:2013dta,PhysRevD.87.076007}, indicate new mechanisms for \CP asymmetries, which should be incorporated in models for future amplitude analyses of charmless three-body $B$ decays.

\section*{Acknowledgements}

\noindent We express our gratitude to our colleagues in the CERN
accelerator departments for the excellent performance of the LHC. We
thank the technical and administrative staff at the LHCb
institutes. We acknowledge support from CERN and from the national
agencies: CAPES, CNPq, FAPERJ and FINEP (Brazil); NSFC (China);
CNRS/IN2P3 and Region Auvergne (France); BMBF, DFG, HGF and MPG
(Germany); SFI (Ireland); INFN (Italy); FOM and NWO (The Netherlands);
SCSR (Poland); MEN/IFA (Romania); MinES, Rosatom, RFBR and NRC
``Kurchatov Institute'' (Russia); MinECo, XuntaGal and GENCAT (Spain);
SNSF and SER (Switzerland); NAS Ukraine (Ukraine); STFC (United
Kingdom); NSF (USA). We also acknowledge the support received from the
ERC under FP7. The Tier1 computing centres are supported by IN2P3
(France), KIT and BMBF (Germany), INFN (Italy), NWO and SURF (The
Netherlands), PIC (Spain), GridPP (United Kingdom). We are thankful
for the computing resources put at our disposal by Yandex LLC
(Russia), as well as to the communities behind the multiple open
source software packages that we depend on.

\addcontentsline{toc}{section}{References}

\providecommand{\href}[2]{#2}

\end{document}